\documentclass[cits]{PoS}

\usepackage{amsmath}

\newcommand{\angav}[1]{\langle #1 \rangle}

\title{A New Dispersive Analysis of $\eta \to 3 \pi$}

\ShortTitle{A New Dispersive Analysis of $\eta \to 3 \pi$}

\author{Gilberto~Colangelo, \speaker{Stefan~Lanz} and Emilie~Passemar\\
        Albert Einstein Center for Fundamental Physics\\
				Institute for Theoretical Physics, University of Bern, Sidlerstrasse 5, 3012 Bern, Switzerland\\
        E-mail: \email{gilberto@itp.unibe.ch}, \email{slanz@itp.unibe.ch}, \email{passemar@itp.unibe.ch}}

\abstract{
We present a new dispersive analysis of the isospin breaking decay $\eta \to 3 \pi$.
The resulting representation of the decay amplitude allows us to determine the quark mass double
ratio $Q$ and we find as a preliminary result $Q = 22.3 \pm 0.4$. Finally, we discuss a
number of improvements that we intend to implement in the future.
}

 \FullConference{\emph{To be published in the proceedings of:}\\
 		6th International Workshop on Chiral Dynamics, CD09\\
 		July 6-10, 2009\\
 		Bern, Switzerland}

\begin{document}

\section{Introduction}

The decay $\eta \to 3 \pi$ is forbidden by isospin symmetry, as Bose statistics does not
allow three pions to form a state with vanishing total isospin and total angular momentum.
Neglecting electromagnetic contributions that are strongly suppressed \cite{Sutherland1966}, the decay amplitude
$A(s,t,u)$ is proportional to $(m_d-m_u)$ or, alternatively, to the quark mass double ratio 
\begin{equation}
	\frac{1}{Q^2} = \frac{m_d^2 - m_u^2}{m_s^2 - \hat{m}^2},
\end{equation}
where $\hat{m} = (m_u+m_d)/2$.
As $\Gamma \propto |A|^2 \propto Q^{-4}$, we can get $Q$ by comparing
a theoretical result for the amplitude with a measurement of the decay
width $\Gamma$. Alternatively, $Q$ can also be calculated from a ratio of meson masses
\cite{Weinberg1977, Kastner+2008}. 

As is well known, the chiral perturbation theory
series converges rather slowly: At tree-level, the decay width is $\Gamma = 66\ \mbox{eV}$
\cite{Cronin1967, Osborn+1970},
while at one-loop it is already $\Gamma = 160\ \mbox{eV}$ \cite{Gasser+1985}, which is still
quite far away from the experimental value $\Gamma = 295\ \mbox{eV}$ \cite{PDG2008}. Of course,
the theoretical values need an input for $(m_d - m_u)$, which is calculated from meson masses:
\begin{equation}
	(m_d - m_u) B_0 = (m_{K^0}^2 - m_{K^+}^2)_{QCD} = m_{K^0}^2 - m_{K^+}^2 - m_{\pi^0}^2 + m_{\pi^+}^2,
\end{equation}
where the second equality relies on Dashen's theorem, stating that at leading order in the low energy
expansion the electromagnetic contribution to the mass differences $m_{K^0}^2 - m_{K^+}^2$ and
$m_{\pi^+}^2 - m_{\pi^0}^2$ is the same \cite{Dashen1969}.
The slow convergence of the chiral series is mainly due to final state rescattering of the pions, which can be very
well treated by means of dispersion relations.

The idea to use dispersion techniques to calculate the $\eta \to 3 \pi$ amplitude is, in fact,
not new: it has been done already more than 10 years ago by Kambor, Wiesendanger and Wyler \cite{Kambor+1996} and by
Anisovich and Leutwyler \cite{Anisovich+1996}. The methods used in these works differ in technical aspects,
but lead to compatible results. We follow the approach as presented in ref.~\cite{Anisovich+1996}.
There has been a lot of activity in this area since then and therefore, we think it is worth to
take a fresh look at this problem. There has been considerable improvement concerning the $\pi \pi$
phase shifts \cite{Ananthanarayan+2001, Colangelo2001, Descotes-Genon+2002, Garcia-Martin+2009}, 
which are the most important physical input in the dispersion relations. Furthermore, there are a number of new measurements
of this decay by the KLOE \cite{Ambrosino+2008,Jacewicz2009}, MAMI \cite{Prakhov+2009,Prakhov2009} and WASA \cite{Adolph+2009,Kupsc2009} 
collaborations, which will be useful for the determination
of the subtraction polynomials, and also a full two-loop calculation in chiral perturbation theory \cite{Bijnens+2007}.

\section{The Method}

We only give a very brief account of the method that we are using and refer to ref.~\cite{Anisovich+1996}
for a more detailed description.
For the charged decay $\eta \to \pi^0 \pi^+ \pi^-$, we define the Mandelstam variables as 
\mbox{$s = (p_{\pi^+} + p_{\pi^-})^2$}, \mbox{$t = (p_{\pi^-} + p_{\pi^0})^2$} and \mbox{$u = (p_{\pi^0} + p_{\pi^+})^2$}.
They are related by $s+t+u = m_\eta^2 + m_{\pi^0}^2 + 2 m_{\pi^+}^2 \equiv 3 s_0$. At the order we
are working the amplitudes for the charged and the neutral channel are related by
\begin{equation}
	A_{\mbox{neutral}}(s,t,u) = A_{\mbox{charged}}(s,t,u) + A_{\mbox{charged}}(t,u,s) + A_{\mbox{charged}}(u,s,t) \;.
\end{equation}
In the following we will restrict ourselves to the charged channel, as we could get the neutral channel easily by the above relation.
We are working in the isospin limit where $m_{\pi^0} = m_{\pi^+}$. It is not a priori clear, what value for the pion mass
in the isospin limit should be used and we choose $m_\pi = m_{\pi^0}$.

Unitarity allows us to relate the imaginary part of the decay amplitude $A_n \dot{=} A_{\eta \to n}$,
where the final state $n$ is some three pion state, to the $\pi \pi$ scattering amplitude and $A_n$
itself as
\begin{equation}
	\mbox{Im}\ A_n = \frac{1}{2} \sum_{n^\prime} \left\{\;  (2 \pi)^4 \delta^4(p_n-p_{n^\prime}) T^*_{n n^\prime} \;\right\} A_{n^\prime} \;.
	\label{eq:ImAn}
\end{equation}
This is a linear constraint for the amplitude and we can thus extract a normalisation factor
\begin{equation}
	A(s,t,u) = - \frac{1}{Q^2} \frac{m_K^2 (m_K^2 - m_\pi^2)}{3 \sqrt{3} m_\pi^2 F_\pi^2}\; M(s,t,u) \;.
\end{equation}
With this choice, the tree-level expression for $M(s,t,u)$ is normalised to 1 at the centre
of the Dalitz plot where \mbox{$s=t=u=s_0$}, and the one-loop amplitude only depends on measurable
quantities. In particular it does not depend on $Q$.

If the discontinuities of $D$- and higher waves are neglected, we can decompose the amplitude into isospin components
in the same way as was done for the $\pi \pi$ scattering amplitude in ref.~\cite{Stern+1993}:
\begin{equation}
	M(s,t,u) = M_0(s) + (s-u) M_1(t) + (s-t) M_1(u) + M_2(t) + M_2(u) - \frac{2}{3} M_2(s) .
	\label{eq:MIsospin}
\end{equation}
As only the S- and P-waves have discontinuities up to two-loop order in chiral perturbation theory, the decomposition
is exact up to that order.

Inserting this decomposition in eq.~\eqref{eq:ImAn}, we get for the discontinuities of the isospin amplitudes
\begin{equation}
	\mbox{disc}\ M_I(s) \; \dot{=} \; \frac{M_I(s+i\epsilon) - M_I(s-i\epsilon)}{2 i}
					= \left\{ M_I(s) + \hat{M}_I(s) \right\} e^{-i \delta_I(s)} \sin \delta_I(s),
	\label{eq:discMI}
\end{equation}
where $\delta_I(s)$ are the $S$- and $P$-wave $\pi \pi$ scattering phase shifts and $I = 0, 1, 2$. The inhomogeneities $\hat{M}_I(s)$
consist of angular averages over the $M_I$:
\begin{subequations}
\begin{align}
	\hat{M}_0(s) &= \frac{2}{3} \angav{M_0} + 2 (s-s_0) \angav{M_1} + \frac{20}{9} \angav{M_2}
									+ \frac{2}{3} \kappa \angav{z M_1}, \\[2mm]
	\hat{M}_1(s) &= \frac{1}{\kappa} \left\{ 3 \angav{z M_0} + \frac{9}{2}(s-s_0)\angav{z M_1}
									- 5 \angav{z M_2} + \frac{3}{2} \kappa \angav{z^2 M_1} \right\}, \\[2mm]
	\hat{M}_2(s) &= \angav{M_0} - \frac{3}{2} (s-s_0)\angav{M_1}+\frac{1}{3}\angav{M_2} - \frac{1}{2} \kappa \angav{z M_1},
\end{align}
	\label{eq:mhat}
\end{subequations}
where
\begin{equation}
	\angav{z^n f}(s) = \frac{1}{2} \int_{-1}^1 dz\ z^n f \left(\frac{3 s_0-s+z\kappa(s)}{2}\right),
	\label{eq:angav}
\end{equation}
and
\begin{equation}
\kappa(s) = \sqrt{\frac{s-4 m_\pi^2}{s}} 
	\sqrt{ \left\{ (m_\eta + m_\pi)^2 - s \right\} \left\{ (m_\eta - m_\pi)^2 - s \right\} } \;.
\end{equation}
The integration variable in eq.~\eqref{eq:angav} is $z = \cos\, \theta$, where $\theta$ is the scattering angle.

These expressions for the discontinuities enable us to write down a set of dispersion integrals, coming
from Cauchy representations of the functions $M_I(s)/\Omega_I(s)$ \cite{Anisovich+1996}:
\begin{subequations}
\begin{align}
	M_0(s) &= \Omega_0(s)\; \left\{ \alpha_0 + \beta_0 s + \gamma_0 s^2 + \frac{s^2}{\pi} \int \limits_{4 m_\pi^2} ^{\infty}
							\frac{ds^\prime}{s^{\prime 2}} \frac{\sin \delta_0(s^\prime) \hat{M}_0(s^\prime)}{|\Omega_0(s^\prime)|
							(s^\prime - s -i \epsilon)} \right\}, \\[3mm]
	M_1(s) &= \Omega_1(s)\; \left\{ \beta_1 s + \frac{s}{\pi} \int \limits_{4 m_\pi^2} ^{\infty} \frac{ds^\prime}{s^\prime} 
							\frac{\sin \delta_1(s^\prime) \hat{M}_1(s^\prime)}{|\Omega_1(s^\prime)| (s^\prime - s -i \epsilon)} \right\}, \\[3mm]
	M_2(s) &= \Omega_2(s)\; \frac{s^2}{\pi} \int \limits_{4 m_\pi^2} ^{\infty} \frac{ds^\prime}{s^{\prime 2}} 
							\frac{\sin \delta_2(s^\prime) \hat{M}_2(s^\prime)}{|\Omega_2(s^\prime)| (s^\prime - s -i \epsilon)}\, .
\end{align}
\label{eq:dispRel}
\end{subequations}
$\Omega_I(s)$ are the so called Omn\`es functions \cite{Omnes1958}, which are the solutions of eq.~\eqref{eq:discMI}
for $\hat{M}_I(s) = 0$ and are given by
\begin{equation}
	\Omega_I(s) = \exp \left\{ \frac{s}{\pi} \int \limits_{4 m_\pi^2}^{\infty} \frac{\delta_I(s^\prime)}{s^\prime (s^\prime - s)}\; ds^\prime \right\}.
\end{equation}

Eqs.~\eqref{eq:dispRel} contain four parameters $\alpha_0$, $\beta_0$, $\gamma_0$ and $\beta_1$.
These subtraction constants are not determined by the dispersion relations and have to be fixed otherwise.
Actually, there are three more of these constants in the equations for $M_1$ and $M_2$, but they can be eliminated
because the decomposition given in eq.~\eqref{eq:MIsospin} is not unique. The $M_I$ can be shifted by a polynomial without
changing the total amplitude $M(s,t,u)$, and the polynomial can be chosen such that three subtraction constants vanish.
The remaining four are determined by a matching to the one-loop result from chiral perturbation theory.

We solve the integral equations numerically by an iterative procedure: We set the $M_I(s)$ to the tree-level result from
chiral perturbation theory and calculate the $\hat{M}_I(s)$ by eqs.~\eqref{eq:mhat} and then get a new result for
the $M_I(s)$ from eqs.~\eqref{eq:dispRel}. The subtraction
constants $\alpha_0$ and $\beta_0$ can now be calculated by a matching to the one-loop amplitude at the Adler zero,
as the amplitude there is protected by the chiral SU(2) $\times$ SU(2) symmetry and does not change much after the one-loop level.
While these two constants have to be determined in every iteration step, the other two, $\gamma_0$ and $\beta_1$, can be
set to their final value right from the start. Once the four constants are fixed, we add the subtraction polynomial and
then repeat the same procedure until the subtraction constants $\alpha_0$ and $\beta_0$ converge.

Special care has to be taken when evaluating the angular averages $\hat{M}_I(s)$ in eqs.~\eqref{eq:mhat}, as the starting and end points
of the integration path, which are functions of $s$, can take complex values. In the first iteration step,
the functions $M_I(s)$ are analytic everywhere and the integral does not depend on the path.
However, the situation changes afterwards: the $M_I(s)$ have a cut along the real axis and the integration
path has to be deformed in such a way that it does not cross the cut \cite{Anisovich+1996a}.

\section{Preliminary Results}

In the future, we will incorporate a number of improvements to the method described in the last section.
The results we present here are therefore only preliminary. The planned extensions are discussed in the next section.

\begin{figure} \begin{center}
		\includegraphics[width=0.49 \textwidth]{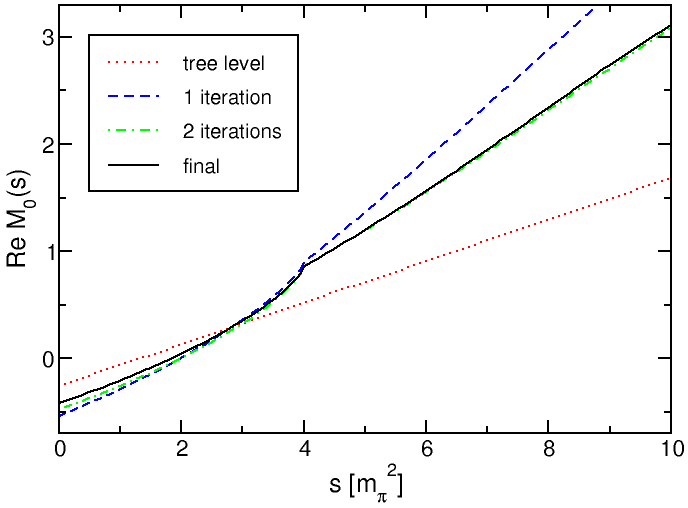}
		\hfill
		\includegraphics[width=0.49 \textwidth]{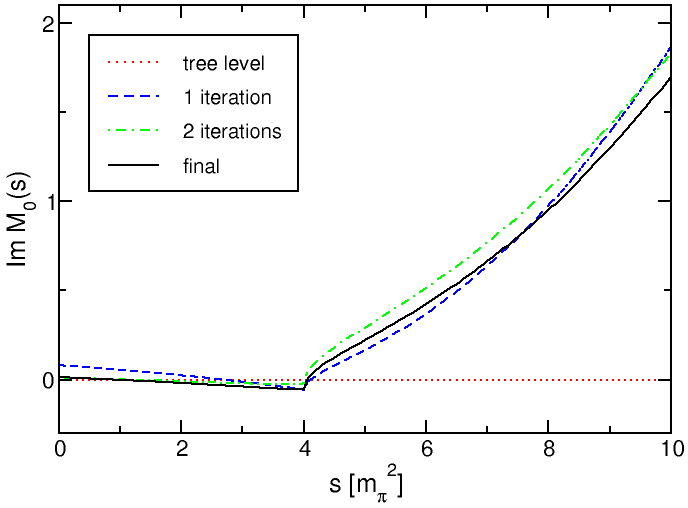}

	\caption{The function $M_0(s)$ and its change during the iterations. The left panel shows the real part, 
						the right one the imaginary part. The red dotted line is the tree-level $\chi$PT
						result used as initial configuration, the blue dashed and the green dash-dotted lines are the results after one and two
						iterations, respectively. The solid black line is the final result after convergence of the subtraction
						constants is reached. In the real part, it is almost impossible to distinguish the result after two iterations
						from the final result.\newline
						To have a smooth representation of $M_0(s)$ near $s = 4 m_\pi^2$, we have to ensure that the threshold
						of the phase shifts agrees with the lower limit of integration. As we know the phase shifts only for
						$m_\pi = m_{\pi^+}$, we also used the charged pion mass to create these figures.}
	\label{fig:M0}
\end{center} \end{figure}

Fig.~\ref{fig:M0} demonstrates how the function $M_0(s)$ develops during the iteration procedure. While for the real part
the curve does not change considerably after the second iteration step, the imaginary part shows that
it is worthwhile going beyond that, as there is still an obvious difference between the result after two iteration
steps and the final result.

\begin{figure} \begin{center}
	\includegraphics[width=0.6 \textwidth]{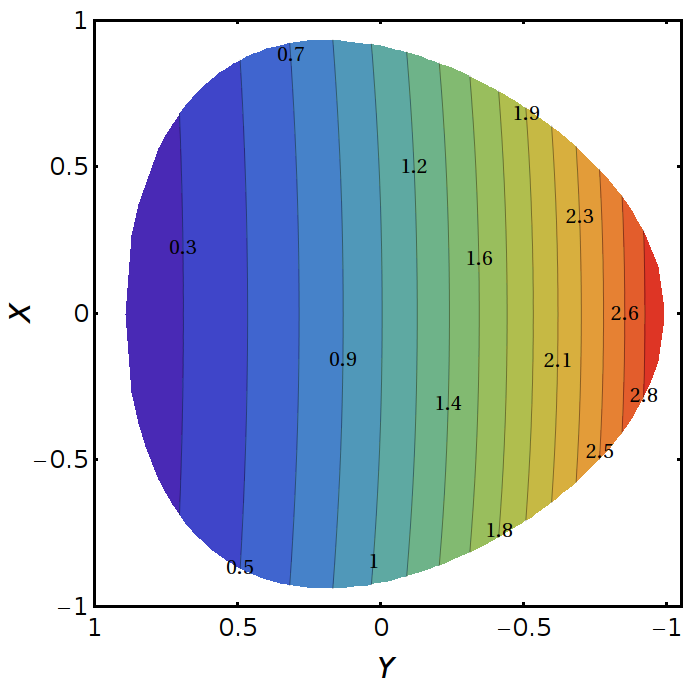}
	\caption{The Dalitz plot for $\eta \to \pi^0 \pi^+ \pi^-$ normalised to 1 for $X=Y=0$.
		The dimensionless Dalitz plot variables $X$ and $Y$ are defined in the text.}
	\label{fig:Dalitz}
 \end{center} \end{figure}

From the functions $M_I(s)$ we can then calculate the amplitude and the Dalitz plot. We use the
dimensionless standard Dalitz plot variables defined as
\begin{equation}
	X = \frac{\sqrt{3}}{2 m_\eta Q_\eta}(u-t), \qquad Y =  \frac{3}{2 m_\eta Q_\eta}\left[ (m_\eta - m_{\pi^0})^2 - s \right] - 1,
	\label{eq:XYdef}
\end{equation}
where
\begin{equation}
	\quad Q_\eta = m_\eta - 2 m_{\pi^+} - m_{\pi^0}.
\end{equation}
Fig.~\ref{fig:Dalitz} shows our result for the Dalitz plot, normalised to 1 for $X=Y=0$.
It agrees with the polynomial representation provided by KLOE \cite{Ambrosino+2008} at the 10 \% level.

\begin{figure} \begin{center}
	\includegraphics[width=0.7 \textwidth]{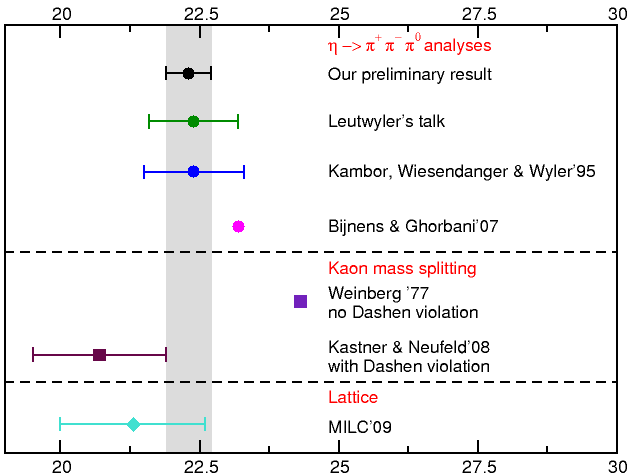}
	\caption{A selection of results for $Q$. Our result is \mbox{$22.3 \pm 0.4$} and is indicated by the grey band. The error
		is only due to the experimental uncertainty on the decay width.
		The other results are taken from Leutwyler's talk at this conference \cite{Leutwyler2009}, from a dispersive
		analysis in ref.~\cite{Kambor+1996}, from a two-loop calculation in $\chi$PT \cite{Bijnens+2007},
		from Weinberg's quark mass ratios \cite{Weinberg1977} and from an analysis including Dashen violation \cite{Kastner+2008}.
		The last value we calculated from the MILC quark mass ratios presented
		by Heller at this conference \cite{Heller2009}.}
	\label{fig:QResults}
\end{center} \end{figure}

With the experimental decay width \mbox{$\Gamma = 295 \pm 20 \ \mbox{eV}$ \cite{PDG2008}}, we get for the quark mass double
ratio \mbox{$Q = 22.3 \pm 0.4$}. This is in good agreement with other results
presented in fig.~\ref{fig:QResults}, which range from 20.7 to 24.3.
The error is only due to the experimental uncertainty in the decay width, as we do not have an estimate for
the theoretical error yet.

\begin{figure} \begin{center}
	\includegraphics[width=0.6 \textwidth]{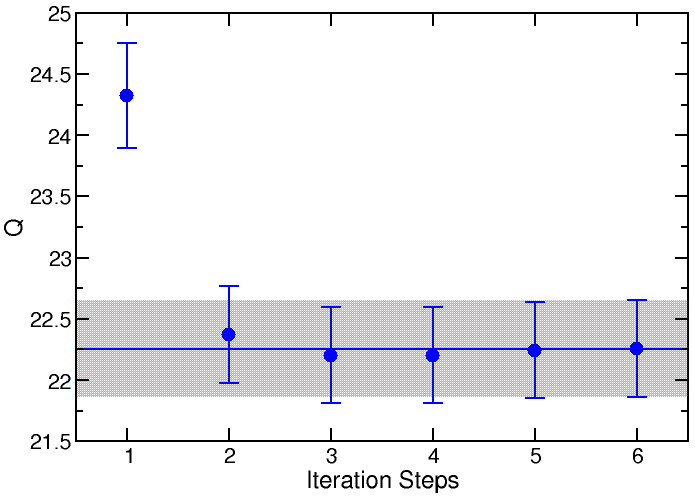}
	\caption{The development of $Q$ over several iteration steps. The blue line is the final result after many iterations,
		the blue band is the corresponding error coming only from the uncertainty of the decay width.}
	\label{fig:qConvergence}
 \end{center} \end{figure}

As a last illustration, fig.~\ref{fig:qConvergence} shows the development of $Q$ over several iteration steps.
Already after two iteration steps, the result for $Q$ is well within the experimental error band of the final result.
However, going beyond that will not only shift the central value within the error bounds, but also reduce the theoretical error,
which is not included in this figure.

\section{Conclusion and Outlook} \label{seq:Conclusion}

We have presented a new dispersive analysis of the decay $\eta \to 3 \pi$ and argued that there have been
several important developments in this and related fields that make a new analysis worthwhile.
A number of preliminary results have been
given, in particular we get for the quark mass double ratio \mbox{$Q = 22.3 \pm 0.4$}, which agrees well with
a number of results from other works. We stress again that this number is preliminary and that the error
given is only due to the uncertainty in the decay width and does not contain any estimate for the
theoretical error.

The method as presented here has a number of shortcomings that we will have to address in the future.
Besides solving these problems, we will also include additional features that have not been
part of earlier dispersive treatments of this decay and that the precision reached in 
experiments nowadays has rendered mandatory.

As mentioned above, we calculate the decay amplitude in the isospin limit and use the mass of the
neutral pion. However, it is not a priori clear, what value for the pion mass should be used and indeed the final
result depends on this choice: setting the pion mass to $m_{\pi^+}$ shifts $Q$ from 22.3 down to 21.0.
The pion mass enters the dispersion integrals in eq.~\eqref{eq:dispRel}
directly, as it is contained in the integration limits and in $\kappa(s)$, and also via the phase shifts.
While it is easy to use the physical masses for the former, it is not for the latter. We would need different
phase shifts for scatterings of pions with different flavours, which are not available at the moment. Indeed the
above-mentioned shift in $Q$ is obtained without altering the pion mass in the phase shifts.

A number of other contributions have not been included yet: electromagnetic interactions \cite{Ditsche+2009,Ditsche2009},
inelasticity and the imaginary parts of D- and higher partial waves. Future extensions of the algorithm will include
estimates of these effects.

The subtraction constants have been determined so far by a matching to the one-loop result from $\chi$PT. We plan to make
use of experimental data instead and calculate the subtraction constants by fitting our Dalitz plot to a measurement.

Of course we can and will also extract a number of other interesting parameters than just $Q$, e.g. the
quark mass ratio $R$, the branching ratio $r = \Gamma_{3 \pi^0} / \Gamma_{\pi^0 \pi^+ \pi^-}$ or the
quadratic slope parameter~$\alpha$.

\acknowledgments

We are grateful to H.~Leutwyler and J.~Gasser for interesting and useful discussions.
The Albert Einstein Center for Fundamental Physics is supported by the ``Innovations- und Kooperationsprojekt
C--13'' of the ``Schweizerische Universit\"atskonferenz SUK/CRUS''. This work was
supported in part by the Swiss National Science Foundation and by EU MRTN--CT--2006--035482 (FLAVIA\textit{net}).

\providecommand{\href}[2]{#2}\begingroup\raggedright\endgroup
 

\begin{thebibliography}{10}

\bibitem{Sutherland1966}
D.~G. Sutherland, {\it Current algebra and the decay $\eta \to 3 \pi$},  {\em
  Phys. Lett.} {\bf 23} (1966) 384.

\bibitem{Weinberg1977}
S.~Weinberg, {\it The problem of mass},  {\em Trans. New York Acad. Sci.} {\bf
  38} (1977) 185--201.

\bibitem{Kastner+2008}
A.~Kastner and H.~Neufeld, {\it The {Kl3} scalar form factors in the standard
  model},  {\em Eur. Phys. J.} {\bf C57} (2008) 541--556,
  [\href{http://xxx.lanl.gov/abs/0805.2222}{{\tt arXiv:0805.2222}}].

\bibitem{Cronin1967}
J.~A. Cronin, {\it Phenomenological model of strong and weak interactions in
  chiral ${U}(3) \times {U}(3)$},  {\em Phys. Rev.} {\bf 161} (1967)
  1483--1494.

\bibitem{Osborn+1970}
H.~Osborn and D.~J. Wallace, {\it $\eta$-${X}$ mixing, $\eta \to 3 \pi$ and
  chiral {L}agrangians},  {\em Nucl. Phys.} {\bf B20} (1970) 23--44.

\bibitem{Gasser+1985}
J.~Gasser and H.~Leutwyler, {\it Chiral perturbation theory: expansions in the
  mass of the strange quark},  {\em Nucl. Phys.} {\bf B250} (1985) 465.

\bibitem{PDG2008}
{\bf Particle Data Group} Collaboration, C.~Amsler {\em et.~al.}, {\it Review
  of particle physics},  {\em Phys. Lett.} {\bf B667} (2008) 1.

\bibitem{Dashen1969}
R.~F. Dashen, {\it Chiral {SU}(3)$\times${SU}(3) as a symmetry of the strong
  interactions},  {\em Phys. Rev.} {\bf 183} (1969) 1245--1260.

\bibitem{Kambor+1996}
J.~Kambor, C.~Wiesendanger, and D.~Wyler, {\it Final state interactions and
  {K}huri-{T}reiman equations in $\eta\to 3\pi$ decays},  {\em Nucl. Phys.}
  {\bf B465} (1996) 215--266,
  [\href{http://xxx.lanl.gov/abs/hep-ph/9509374}{{\tt hep-ph/9509374}}].

\bibitem{Anisovich+1996}
A.~V. Anisovich and H.~Leutwyler, {\it Dispersive analysis of the decay $\eta
  \to 3 \pi$},  {\em Phys. Lett.} {\bf B375} (1996) 335--342,
  [\href{http://xxx.lanl.gov/abs/hep-ph/9601237}{{\tt hep-ph/9601237}}].

\bibitem{Ananthanarayan+2001}
B.~Ananthanarayan, G.~Colangelo, J.~Gasser, and H.~Leutwyler, {\it Roy equation
  analysis of $\pi \pi$ scattering},  {\em Phys. Rept.} {\bf 353} (2001)
  207--279, [\href{http://xxx.lanl.gov/abs/hep-ph/0005297}{{\tt
  hep-ph/0005297}}].

\bibitem{Colangelo2001}
G.~Colangelo, J.~Gasser, and H.~Leutwyler, {\it $\pi \pi$ scattering},  {\em
  Nucl. Phys.} {\bf B603} (2001) 125--179,
  [\href{http://xxx.lanl.gov/abs/hep-ph/0103088}{{\tt hep-ph/0103088}}].

\bibitem{Descotes-Genon+2002}
S.~Descotes-Genon, N.~H. Fuchs, L.~Girlanda, and J.~Stern, {\it Analysis and
  interpretation of new low-energy $\pi \pi$ scattering data},  {\em Eur. Phys.
  J.} {\bf C24} (2002) 469--483,
  [\href{http://xxx.lanl.gov/abs/hep-ph/0112088}{{\tt hep-ph/0112088}}].

\bibitem{Garcia-Martin+2009}
R.~Garcia-Martin, R.~Kaminski, J.~R. Pelaez, and F.~J. Yndurain, {\it Once
  subtracted {R}oy-like dispersion relations and a precise analysis of $\pi
  \pi$ scattering data},  {\em \emph{in proceedings of} International Workshop
  on Effective Field Theories,} {\bf \pos{PoS(EFT09)052}} (2009).

\bibitem{Ambrosino+2008}
{\bf KLOE} Collaboration, F.~Ambrosino {\em et.~al.}, {\it Determination of
  $\eta\to\pi^+\pi^-\pi^0$ {D}alitz plot slopes and asymmetries with the {KLOE}
  detector},  {\em JHEP} {\bf 05} (2008) 006,
  [\href{http://xxx.lanl.gov/abs/0801.2642}{{\tt arXiv:0801.2642}}].

\bibitem{Jacewicz2009}
M.~Jacewicz, {\it $\eta$/$\eta^\prime$ physics at {KLOE}},  {\em \emph{in
  proceedings of} 6th International Workshop on Chiral Dynamics,} {\bf
  \pos{PoS(CD09)045}} (2009).

\bibitem{Prakhov+2009}
{\bf Crystal Ball at MAMI} Collaboration, S.~Prakhov {\em et.~al.}, {\it
  Measurement of the slope parameter $\alpha$ for the $\eta\to 3\pi^0$ decay
  with the {C}rystal {B}all at {MAMI-C}},  {\em Phys. Rev.} {\bf C79} (2009)
  035204, [\href{http://xxx.lanl.gov/abs/0812.1999}{{\tt arXiv:0812.1999}}].

\bibitem{Prakhov2009}
S.~Prakhov, {\it Study of the $\eta \to 3 \pi^0$ decay with the {C}rystal
  {B}all at {MAMI-C}},  {\em \emph{in proceedings of} 6th International
  Workshop on Chiral Dynamics,} {\bf \pos{PoS(CD09)044}} (2009).

\bibitem{Adolph+2009}
{\bf WASA-at-COSY} Collaboration, C.~Adolph {\em et.~al.}, {\it Measurement of
  the $\eta \to 3 \pi^0$ {D}alitz plot distribution with the {WASA} detector at
  {COSY}},  {\em Phys. Lett.} {\bf B677} (2009) 24--29,
  [\href{http://xxx.lanl.gov/abs/0811.2763}{{\tt arXiv:0811.2763}}].

\bibitem{Kupsc2009}
A.~Kupsc, {\it Studies of eta meson decays with {WASA}},  {\em \emph{in
  proceedings of} 6th International Workshop on Chiral Dynamics,} {\bf
  \pos{PoS(CD09)046}} (2009).

\bibitem{Bijnens+2007}
J.~Bijnens and K.~Ghorbani, {\it $\eta \to 3 \pi$ at two loops in chiral
  perturbation theory},  {\em JHEP} {\bf 11} (2007) 030,
  [\href{http://xxx.lanl.gov/abs/0709.0230}{{\tt arXiv:0709.0230}}].

\bibitem{Stern+1993}
J.~Stern, H.~Sazdjian, and N.~H. Fuchs, {\it What $\pi$-$\pi$ scattering tells
  us about chiral perturbation theory},  {\em Phys. Rev.} {\bf D47} (1993)
  3814--3838, [\href{http://xxx.lanl.gov/abs/hep-ph/9301244}{{\tt
  hep-ph/9301244}}].

\bibitem{Omnes1958}
R.~Omn\`es, {\it On the solution of certain singular integral equations of
  quantum field theory},  {\em Nuovo Cim.} {\bf 8} (1958) 316--326.

\bibitem{Anisovich+1996a}
V.~V. Anisovich, L.~G. Dakhno, and V.~A. Nikonov, {\it {QCD-Motivated Pomeron
  and High Energy Hadronic Diffractive Cross Sections}},  {\em Phys. Atom.
  Nucl.} {\bf 59} (1996) 702--708,
  [\href{http://xxx.lanl.gov/abs/hep-ph/9511211}{{\tt hep-ph/9511211}}].
  [\emph{Yad. Fiz.} \textbf{59N4} (1996) 735--741].

\bibitem{Leutwyler2009}
H.~Leutwyler, {\it Light quark masses},  {\em \emph{in proceedings of} 6th
  International Workshop on Chiral Dynamics,} {\bf \pos{PoS(CD09)005}} (2009).

\bibitem{Heller2009}
U.~M. Heller, {\it {MILC} results for light pseudoscalars},  {\em \emph{in
  proceedings of} 6th International Workshop on Chiral Dynamics,} {\bf
  \pos{PoS(CD09)007}} (2009).

\bibitem{Ditsche+2009}
C.~Ditsche, B.~Kubis, and U.-G. Meissner, {\it Electromagnetic corrections in
  $\eta \to 3 \pi$ decays},  {\em Eur. Phys. J.} {\bf C60} (2009) 83--105,
  [\href{http://xxx.lanl.gov/abs/0812.0344}{{\tt arXiv:0812.0344}}].

\bibitem{Ditsche2009}
C.~Ditsche, {\it Electromagnetic corrections in $\eta \to 3 \pi$ decays},  {\em
  \emph{in proceedings of} 6th International Workshop on Chiral Dynamics,} {\bf
  \pos{PoS(CD09)043}} (2009).

\end{thebibliography}
\end{document}